\pdfoutput=1
\documentclass[journal]{IEEEtran}
\usepackage{ifpdf}
\usepackage{amsmath, amssymb}

\ifCLASSINFOpdf
  \usepackage[pdftex]{graphicx}
\else
  \usepackage[dvips]{graphicx}
\fi
\usepackage{subcaption}
\usepackage{lipsum}

\begin{document}
\title{Massive MIMO testbed - Implementation and Initial Results in System Model Validation}

\author{Changchun~Zhang,~\IEEEmembership{Student Member,~IEEE,}
        Robert~C.~Qiu,~\IEEEmembership{Fellow,~IEEE,}
        %and~Jane~Doe,~\IEEEmembership{Life~Fellow,~IEEE}% <-this % stops a space
%\thanks{M. Shell is with the Department
%of Electrical and Computer Engineering, Georgia Institute of Technology, Atlanta,
%\thanks{Manuscript received April 19, 2005; revised December 27, 2012.}
}

\maketitle

\begin{abstract}
This paper presents the design and implementation of a novel SDR based massive MIMO testbed with up to 70 nodes 
built at Tennessee Technological University. The deployment can reach a $30 \times 30$ antenna MIMO scheme.
With this testbed, we are able to measure the channel matrix and compute the achievable rate of the massive MIMO system using experimental data. 
The measured channel capacity is linearly increasing with the number of antennas of the base station. We also demonstrate the channel reciprocity
including the circuits impact from the transmitter and receiver. We show that the Vandermonde channel model is more realistic to describe the massive MIMO architecture than the widely used Gaussian channel model, in terms of capacity.
By adjusting the range for angle of arrival $\alpha$ and the base station antenna distance $d$ during the simulation, we find out the Vandermonde model agrees with our measured capacity
at a certain $\alpha$ for each selected $d$ and the $\alpha$ is very close to that of the experiment deployment. It is the first time that the feasibility of Vandermonde channel model is demonstrated by the experiment for massive MIMO. 
\end{abstract}


% \fi
\IEEEpeerreviewmaketitle

\section{introduction}
%literature review...
%Massive MIMO is the working horse of the 5G communications system.
%The theoretical research demands the real testbed to ...
%Existing MIMO testbeds are not big enough ...
%Gaussian model dominates the research model in Massive MIMO but not realistic.
%Vandermonde channel model is more close to the real system ....

Massive MIMO, as an candidate for one of the disruptive technologies of the next generation (5G) communications system, promises significant gains in wireless data rates and link reliability ~\cite{rusek2013scaling} ~\cite{boccardi2013five}. By using large numbers of antennas at the base station, massive MIMO helps to focus the radiated energy toward the intended direction
while minimizing the intra and intercell interference.

While promising, the research challenges of massive MIMO also emerge with the dramatically increased number of antennas at the base station. Obtaining channel state information is an essential operation for communications but 
the pilot contamination could occur due to the limited pilots for massive MIMO. 

Most of the current theoretical analysis on large MIMO assume a Gaussian Channel matrix ~\cite{wagner2012large} ~\cite{couillet2011deterministic}. However, Gaussian channel could be not accurate in real environment.
The channel could be correlated due to spatial impact. Specially, in the sectorized celluar network outdoors, signals from the mobile users in a sector could be distributed within a narrow angle ]. The Vandermonde matrix
can be used to depict the channel model of the outdoor LOS propagation path for the large antennas system ~\cite{ryan2008random} ~\cite{debbah2008random} ~\cite{tucci2011eigenvalue}. Most of the current research on massive-MIMO is limited to theory based on the analytical and simulation results,
and lacks of the experimental support. Thus our experimental research is motivated to verify the theoretical results and to find the effect of massive MIMO in real propagation environment. We built a prototype of massive MIMO  of testbed with the commercial SDR platfrom - USRP N210. With $30$ high performance computer and $70$ SDR nodes, we developed a massive MIMO experiment testbed supporting up to $30 \times 30$ scheme. We also developed the software to support the distributed data collection and computing architecture. Our preliminary experimental work based on this testbed include 
\begin{itemize}
	\item Implement of the TDMA protocol to ensure the orthogonality in channel measurement;
	\item Compute the massive MIMO capacity based on the measured channel matrix, pre-coding is also considered;
\end{itemize}
Our experimental results show the linearity of the capacity with the increasing number of antennas by 
assuming it is the same with the number of mobile users. Further, more importantly, we found out that our measured capacity agrees with the simulated capacity under Vandermonde channel model.   It is the first time that 
the feasibility of Vandermonde channel model is demonstrated by experimental result. 

tntech\section{Theoretical Model}

\begin{figure}
\centering
\includegraphics[width=3.4in]{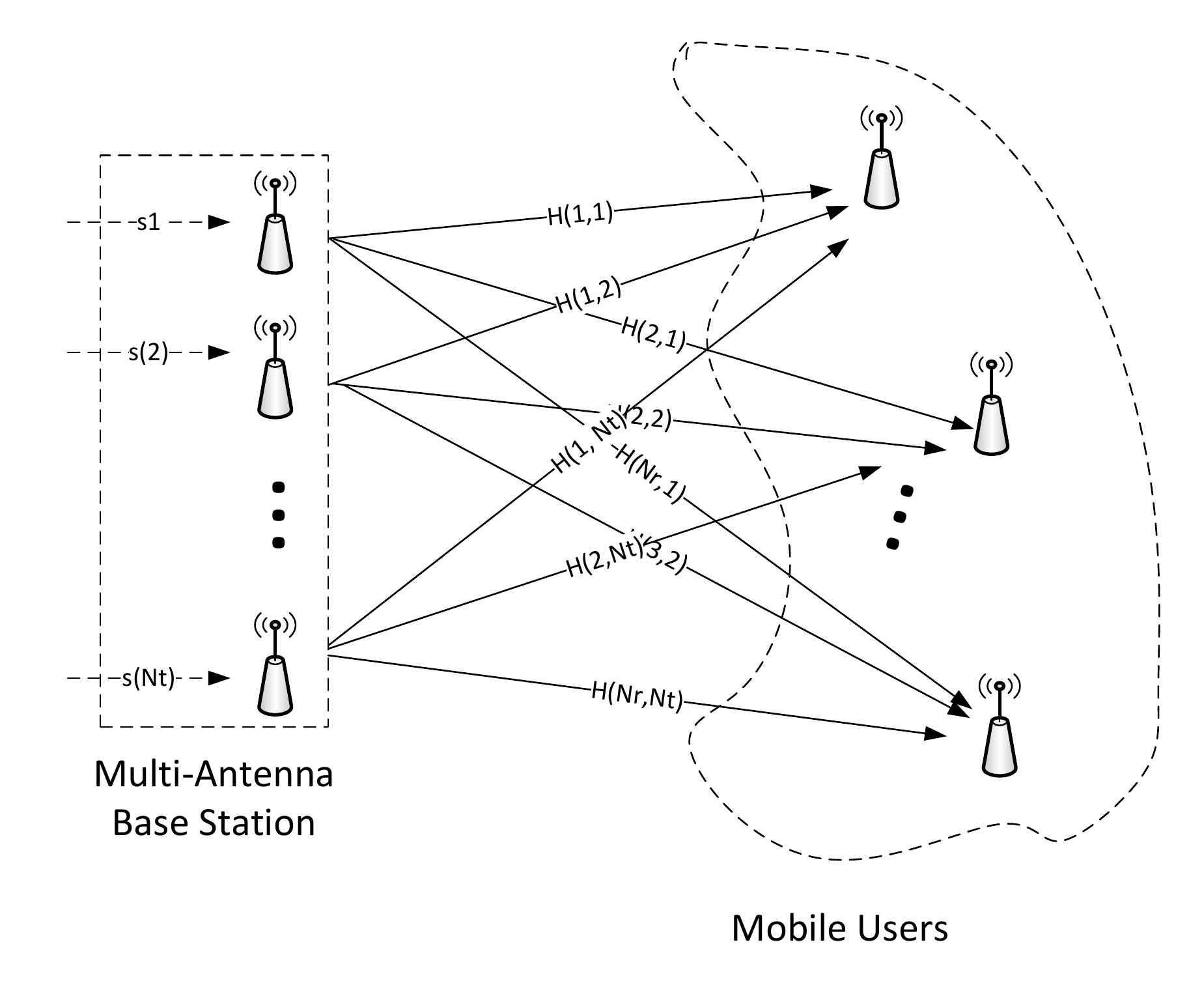}
\caption{System model of MU-MIMO system with ${N_t}$-antenna base station and ${N_r}$ mobile users}
\label{mu_mimo_sys}
\end{figure}

Considering a point-to-point MIMO system in which there are ${N_t}$ transmit antennas and ${N_r}$ receive antennas. 
For the narrow band channel, assuming the transmitters transmit the data simultaneously, the mathematical
description of the signal model is as below:

\begin{equation}
{\mathbf{y}} = \sqrt \rho  {\mathbf{Hs}} + {\mathbf{n}}
\end{equation}

where ${\mathbf{s}} \in {\mathbb{C}^{{N_t}}}$ is the transmitted vector satisfying the ${\rm E}\left[ {{\bf{s}}{{\bf{s}}^{\rm{H}}}} \right] = 1$,
$\rho $ is the common SNR(signal-to-noise-ratio) at each receive antenna and $\bf{n}$ is a vector of complex Guassian noise complying
to zero mean and unit-variance i.i.d(independent and identical distribution), ${\mathbf{H}} \in {\mathbb{C}^{{N_r} \times {N_t}}}$ is the channel
response matrix.

\subsection{Capacity and achievable rates} 

Under the assumption that the receiver has perfect knowledge of channel matrix ${\mathbf{H}}$, the capacity of the ${N_t} \times {N_r}$ MIMO 
channel is computed by 

\begin{equation}
C = {\log _2}\det \left( {{{\text{I}}_{{N_r}}} + \frac{\rho }{{{N_t}}}{\mathbf{H}}{{\mathbf{H}}^H}} \right)
\end{equation}
where ${{{\text{I}}_{{N_r}}}}$ is the identity matrix and the $H$ means hermitian transposition.

\subsection{Vandermonde Channel Model}
\label{vand_mode}
In ~\cite{ryan2008random} ~\cite{debbah2008random} ~\cite{tucci2011eigenvalue}
the Vandermonde random matrix 
\begin{equation}
\frac{1}{{\sqrt N }}\left[ {\begin{array}{*{20}{c}}
  1& \cdots &1 \\ 
  {{e^{ - j2\pi \frac{d}{\lambda }\sin \left( {{\theta _1}} \right)}}}& \cdots &{{e^{ - j2\pi \frac{d}{\lambda }\sin \left( {{\theta _N}} \right)}}} \\ 
   \vdots & \ddots & \vdots  \\ 
  {{e^{ - j2\pi \left( {M - 1} \right)\frac{d}{\lambda }\sin \left( {{\theta _1}} \right)}}}& \cdots &{{e^{ - j2\pi \left( {M - 1} \right)\frac{d}{\lambda }\sin \left( {{\theta _N}} \right)}}} 
\end{array}} \right]
\end{equation}
is introduced to describe the channel model for a base station receiver with $M$ antennas and $N$ mobiles, where $d$ is the antenna spacing and $\lambda$ is the wavelength.
As in fig.~\ref{mu_mimo_sys}, the angles of arrival are supposed to be uniformed distributed within $\left( { - \alpha ,\alpha } \right)$. The elements of the Vandermonde matrix can also have phases with uniform distribution for comparison.
The received signal at the base station is given by 
\begin{equation}
{\mathbf{y}} = {\mathbf{V}}{{\mathbf{P}}^{\frac{1}{2}}}{\mathbf{s}} + {\mathbf{n}}
\end{equation}
where ${\mathbf{y}}$, $\mathbf{s}$, $\mathbf{n}$ are respectively the $M \times 1$ received vector, the $N \times 1$  transmit vector, and the 
$M \times 1$ additive noise, ${\mathbf{V}}$ is the Vandermonde channel matrix, ${\mathbf{P}}$ is the power gain matrix which can be set as identity matrix in simulation. 

The Vandermonde model, compared with Gaussian channel model, is close to the real massive MIMO system from the architecture point of view.  

%\subsection{Pre-coding over the measured known channel}
%With the known down-link channel matrix of the MU-MIMO system, the linear pre-coding can be applied to the transmitted symbols vector $s$.
%Although generally suboptimal, the zero-forcing pre-coding scheme with optimal transmitting power allocation can achieve the same
%asymptotic sum capacity as the dirty paper coding (DPC). 
%
%Denoting ${\bf{x}}$ as the transmitting vector after pre-coding with power constraint $E\left( {{{\left\| {\bf{x}} \right\|}^2}} \right) \le 1$, we have
%\begin{equation}
%\label{precode}
%{\bf{x}} = {\bf{Us}}
%\end{equation}
%while ${\bf{U}} = {\bf{W}}\sqrt {\bf{P}} $ and $\bf{W}$ is the matrix of pre-coding algorithm, ${\bf{P}} = diag\left\{ {{P_1}, \cdots ,{P_{{N_r}}}} \right\}$
 %is the transmitting power scaling factor matrix.
%For zero-forcing pre-coding, 
%\begin{equation}
%\label{zf_w}
%{\bf{W}} = {{\bf{H}}^H}{\left( {{\bf{H}}{{\bf{H}}^H}} \right)^{ - 1}}
%\end{equation}
%and the achievable sum rate becomes 
%\begin{equation}
%\label{czf}
%{C_{ZF}} = \mathop {\max }\limits_{{P_i}:\sum\limits_{i = 1}^{{N_r}} {{\gamma ^{ - 1}}{P_i}}  \le 1} \sum\limits_{i = 1}^{{N_r}} {\log \left( {1 + {P_i}} \right)}
%\end{equation} 
%where ${\gamma _i} = \frac{1}{{{{\left\| {{{\bf{w}}_i}} \right\|}^2}}}$ and ${{\mathbf{w}}_i} \in {\mathbb{C}^{{N_t} \times 1}}$ is the column vector of matrix 
%${\mathbf{W}}$.
%The optimal ${{P_i}}$ can be found by waterfilling.

\section{System design and experiment setup}

\subsection{Channel Measurement}
\label{chan_measure}

The channel matrix is key to compute the MIMO system capacity. Denote ${h_{ij}}$ as the frequency response for the link between the transmit antenna $i$ 
and receive antenna $j$. Assuming the narrowband channel, the channel can be measured by the PN sequence correlation ~\cite{islam2013wireless}.

At the transmitter, a PN sequence with length of 512, ${\mathbf{x}} = [{c_0}, \cdots {c_i}, \cdots {c_{510}}]$, is transmitted. The auto-correlation function of
the PN sequence is 
\begin{equation}
{R_{xx}}\left( n \right) = \left\{ {\begin{array}{*{20}{c}}
  1&{n = 0,N, - N,2N, - 2N, \cdots } \\ 
  { - \frac{1}{N}}&{otherwise} 
\end{array}} \right.
\end{equation}

At the receiver, the correlation between the received sequence, ${\mathbf{y}}$ and the known transmitted PN sequence, ${\mathbf{x}}$ is
\begin{equation}
\label{corr}
{R_{xy}}\left( n \right) = {h_{ij}}{R_{xx}}\left( n \right)
\end{equation}
Thus the $h_{ij}$ is computed based on eq.~\ref{corr}.

In the real channel measurement, the TDD(Time division duplex) is introduced to measure the channel matrix by ensuring the orthogonality of the channel sounding signal. 

%\subsection{Channel Reciprocity} \\
%\label{chan_recip}
%
%For the antenna $i$ and $j$, if the uplink and downlink work in TDD mode, the channel reciprocity will be useful for the precoding in MIMO system.
%Channel reciprocity means the $h_{ij} = h_{ji}$ if they only represent the air channels. However, in the real channel measurement, the channel state information obtained from above 
%subsection also includes the impact of the circuits. 
%
%Given the $h$ is the air channel between antenna $i$ and $j$, the measured channel $h_{ij}$ and $h_{ji}$ follow the model depicted as Fig.~\ref{reciprocity}, where
%where $T\left( i \right)$, ${R\left( j \right)}$, $R\left( i \right)$, $T\left( j \right)$ represent the effect from circuits like upper/down conversion, filters, etc., for
%both the upper and down links.
%\begin{figure}
%\centering
%\includegraphics[width=3.6in]{figs/resip.pdf}
%\caption{Reciprocity mode for TDD channel}
%\label{reciprocity}
%\end{figure}
%
%Thus we have
%\begin{equation}
%\begin{gathered}
  %{h_{ij}} = T\left( i \right) \cdot h \cdot R\left( j \right) \hfill \\
  %{h_{ij}} = R\left( i \right) \cdot h \cdot T\left( j \right) \hfill \\ 
%\end{gathered} 
%\end{equation}
%
%Usually, the relative calibration is sufficient for the precoding as we have
%\begin{equation}
%\frac{{{h_{ij}}}}{{{h_{ji}}}} = \frac{{T\left( i \right) \cdot R\left( j \right)}}{{R\left( i \right) \cdot T\left( j \right)}}
%\end{equation}
%which is constant in ideal situation.

\subsection{Experiment Deployment and System Architecture}
\begin{figure}
\centering
\includegraphics[width=3.6in]{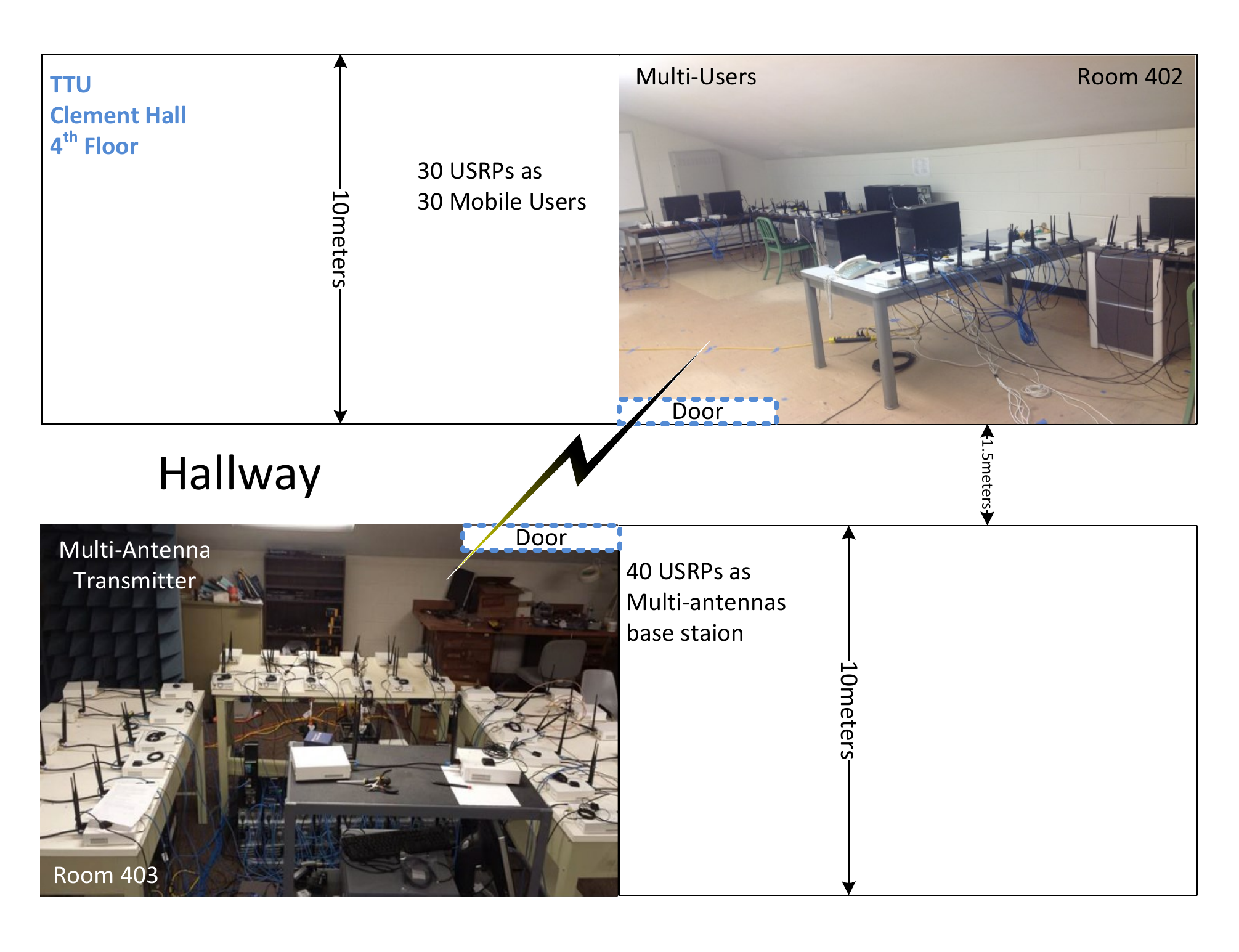}
\caption{The experiment deployment of the MU-MIMO system.}
\label{deployment}
\end{figure}

In the wireless networking systems lab (WNSL) of Tennessee Tech university (TTU), we built a massive MIMO testbed based the general 
SDR platform - USRP. Seventy USRPs controlled by thirty high performance PCs are deployed on the 4th floor of Clement Hall on the 
campus of TTU. As shown in Fig.~\ref{deployment}, two groups of USRPs are placed in two separate rooms. Forty USRPs are deployed 
in the WNSL to emulate multi-antenna base station, while the rest of the USRPs are randomly arranged in another room
acting as mobile users with single antenna.

\begin{figure}
\centering
\includegraphics[width=3.6in]{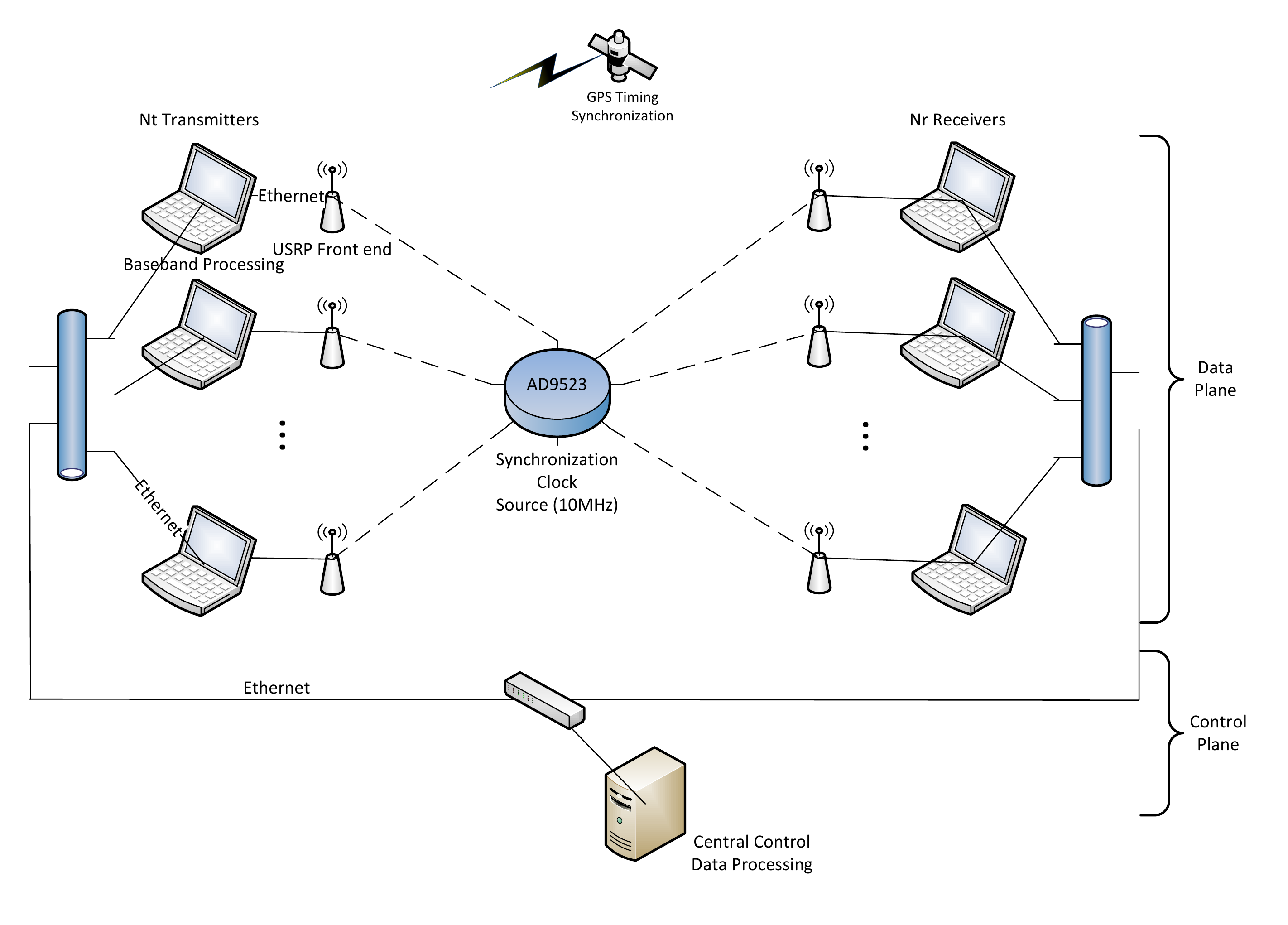}
\caption{The system architecture of the massive MIMO testbed over USRP SDR.}
\label{system_arch}
\end{figure}

The massive MIMO system is composed of data plane and control plane, as depicted by Fig.~\ref{system_arch}. The data plane is 
responsible for the channel measurement and waveform data processing. On the data plane, 30 PCs are configured as baseband processing units.
Every baseband processing unit is connected to arbitrary number of USRP front-ends acting as either transmitter or receiver, through Ethernet. 
On the control plane, one central control PC is connected with
all the baseband processing units to control the procedures of the experiment including the start and stop of measurement, parameters configuration, etc.
The communications between the baseband processing unit and control unit is also over Ethernet and the socket/TCP programming.

%In the following subsections, we present the detail design of the testbed from several key aspects.

\subsection{Massive MIMO testbed over time division protocol}
The channel matrix measurement is a challenging task that is achieved by our massive MIMO testbed. Firstly, we want to measure the data within the time frame that the channel 
can be regarded as static. Secondly, there are so many links ($30 \times 30$) to be measured within this time frame. The orthogonality needs to be ensured 
that the measured result for each link is not interfered by all the rest of the links. 
We choose the time division method to separate the channel measurements for all the links.  
Under the time division strategy, our massive MIMO testbed is equivalent to a TDMA network. $30$ transmitters only send the PN sequence within their own time slots, while the mobile users can be set as receive only mode.
 
 During the experiment, suppose the number of the transmitters is the same with the number of mobile
users, so we actually deployed a $30 \times 30$ MIMO system which is regarded as a massive MIMO testbed as the scale is much bigger than the current commercial MIMO system.

%By applying the channel reciprocity, we can get channel information for both the uplink and downlink by only
%measure the uplink information. Meanwhile, for simplicity, we can also get the both the uplink and downlink information within one round of test, by configuring all 
%the USRPs nodes as TDMA nodes, whether they are emulating base station antenna or mobile users.

\subsection{Synchronization}
The timing synchronization for all the TDMA nodes is achieved 
by the GPS signal as shown in Fig.~\ref{system_arch}. The timing precision provided by the GPS signal reaches nano-second, which is sufficient for the time slot scheduling.
Every USRP is equipped with a dedicated GPS antenna that can feed the timing information to the front end. The baseband samples sent from the PC to the USRP can be arranged
within the specified time frame for transmitting by adding the time tag information at the start of sample series. 

We also provide the clock synchronization for all the USRP
front-ends by the clock distributing board AD9523 as in figure.~\ref{system_arch}. The common clock 
between the transmitters and receivers helps to remove the carrier frequent offset(CFO) that is introduced by different clock sources. 

\section{Experiment results}
This massive MIMO testbed can be scaled to at most $30 \times 30$. The measured channel matrix is used to calculate the real MIMO system capacity. 

%Figure.~\ref{tdma_wave} gives an example of TDMA waveform to measure the $4 \tims 4$. 
In the experiment, if we set the time slot 
duration to $50ms$, we can get a round of TDMA data by $200ms$ for $4 \time 4$ MIMO system. $30 \times 30$ configuration costs $1.5s$
to get a round of data. Basically the channel can be regarded as static within several seconds for our current static deployment because there no other impact during the experiment.

\begin{figure*}
        \centering
        \begin{subfigure}[h]{0.5\textwidth}
         \centering
							\includegraphics[width=3.4in]{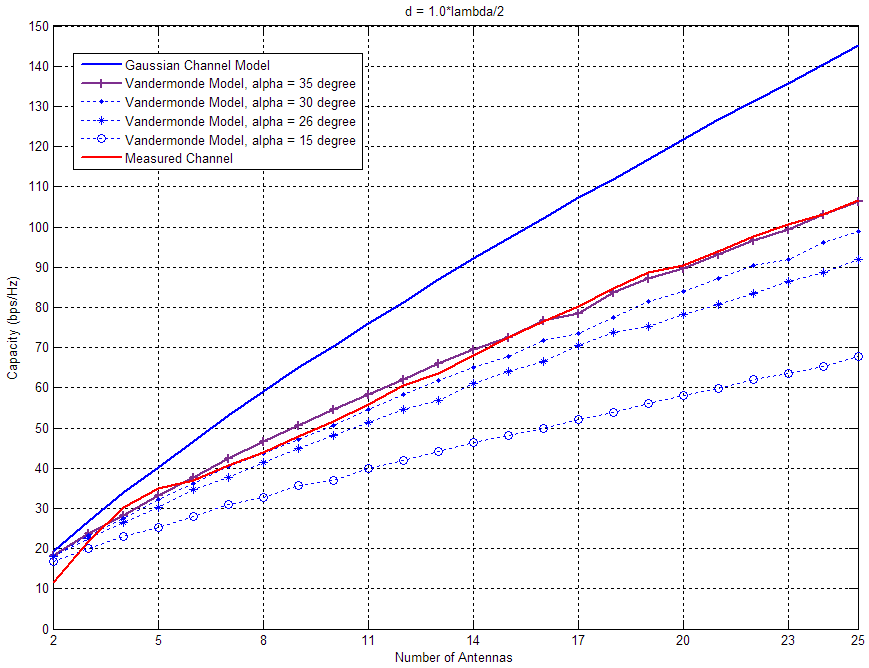}
                \caption{$d=1.0\lambda /2$}
        \end{subfigure}%
        ~ %add desired spacing between images, e. g. ~, \quad, \qquad, \hfill etc.
        \begin{subfigure}[h]{0.5\textwidth}
				\centering
                \includegraphics[width=3.4in]{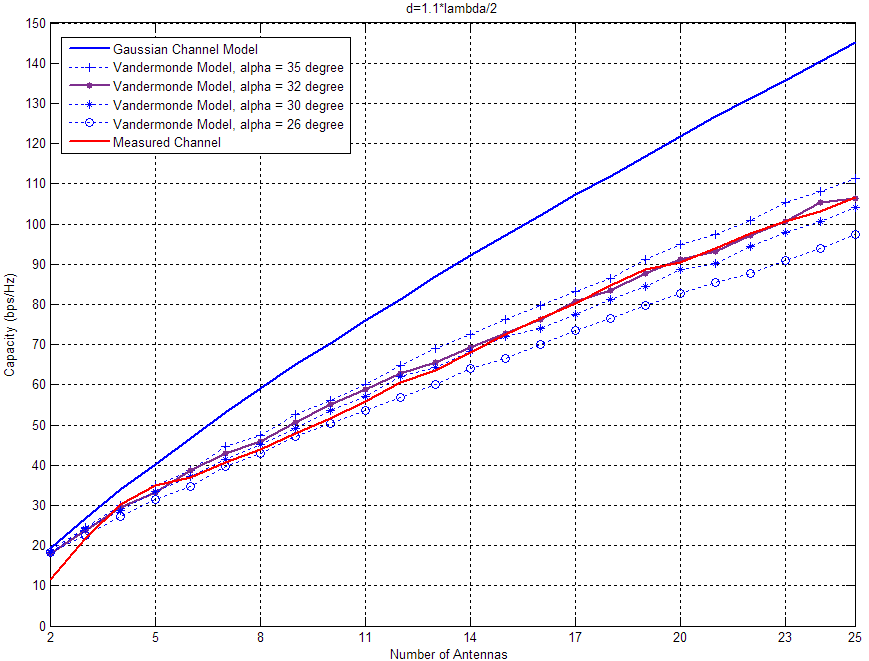}
                \caption{$d=1.1\lambda /2$}
        \end{subfigure}
				
				\begin{subfigure}[h]{0.5\textwidth}
				\centering
                \includegraphics[width=3.4in]{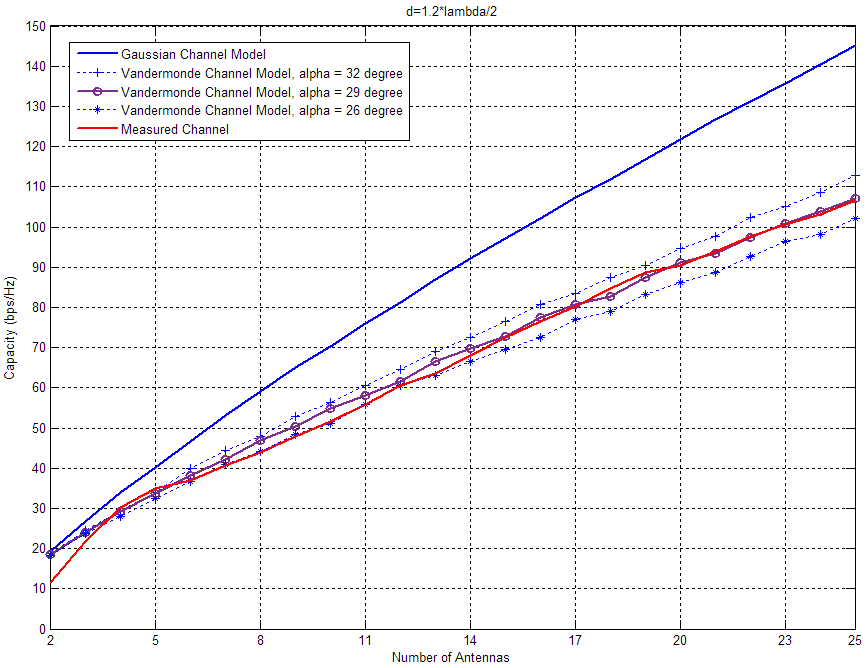}
                \caption{$d=1.2\lambda /2$}
        \end{subfigure}%
        ~ %add desired spacing between images, e. g. ~, \quad, \qquad, \hfill etc.
        \begin{subfigure}[h]{0.5\textwidth}
				\centering
                \includegraphics[width=3.4in]{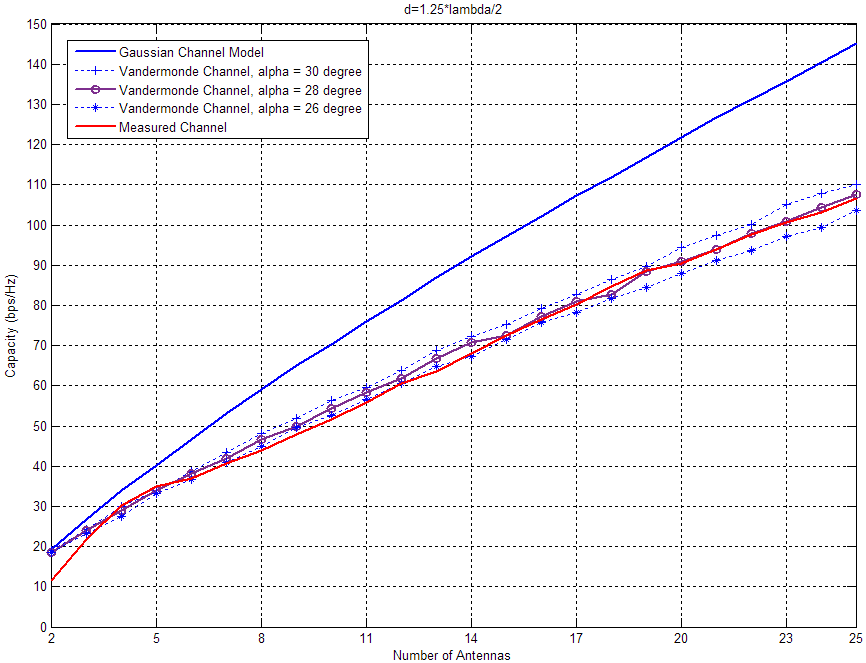}
                \caption{$d=1.25\lambda /2$}
        \end{subfigure}
        \caption{Comparison of measured channel with Vandermonde channel model by adjusting $d$ and $\alpha$,
				         in terms of capacity.}
				\label{cap_vand}
\end{figure*}

We did the simulations for the capacity under the Vandermonde channel model, by adjusting two parameters $d$ and $\alpha$ based on section \ref{vand_mode}. Fig.~\ref{cap_vand} shows the experimental results compared with the simulations. The experimental frequency is 926MHz, the corresponding wavelength is $\lambda=32.4cm$. The average distance between the adjacent nodes for base station is about $20cm$. We selected the parameter $d$ in $\left \{  1.0\lambda/2, 1.1\lambda/2, 1.2\lambda/2, 1.25\lambda/2\right \}$ and did the simulation for Vandermonde model by adjusting $\alpha$ for each fixed $d$.

Firstly, we demonstrate the linearity of the capacity with the number of antennas by the measured channel capacity in Fig.~\ref{cap_vand}. The capacity of the system increases linearly with the number of antennas, for both the simulated Gaussian channels, Vandermonde channels and the measured channels. 
%
%\begin{figure}
%\centering
%\includegraphics[width=3.6in]{figs/capacity.png}
%\caption{Capacity linearity over number of antennas}
%\label{capacity}
%\end{figure}

%\begin{figure}
%\centering
%\includegraphics[width=3.6in]{figs/tdma_wave.png}
%\caption{TDMA timeslot waveform, 4 nodes as example }
%\label{tdma_wave}
%\end{figure}

Secondly, we found the feasibility to fit the massive MIMO channel model to the Vandermonde channel model, by the comparison of capacities between the real channel  measurement and the Vandermonde channel model.
%In our experimental deployment as Fig.~\ref{deployment}, the range for angled of arrival is round in $(-{26 ^\circ}, {26 ^\circ})$.
%We give the simulation curves for Vandermonde channel model with 
%$\alpha  = {40^ \circ },{35^ \circ },{30^ \circ },{26^ \circ },{15^ \circ }$ respectively.
Fig.~\ref{cap_vand} shows observations including :
\begin{itemize}
\item The capacity of Vandermonde channel decreases when $\alpha$ is getting smaller for fixed $d$.
\item The capacity of Vandermonde channel increases when $d$ is bigger for fixed $\alpha$.
\item For every fixed $d$, we can find an $\alpha$ to fit the Vandermonde model to measured channel in capacity with 
minimum mean error (MME). Table.~\ref{mme_tab} shows the MME value for different $\alpha$ under each fixed $d$. Bold $\alpha$ represents the optimal $\alpha$ for each $d$. 
\end{itemize}
We get insight from above that the Vandermonde channel model is more realistic than the Gaussian channel model in depicting the massive MIMO system. The Gaussian model does not reflect the topology of the real MIMO system while the Vandermonde model is more illustrative as the parameter $\alpha$ and $d$ are contained. Also, the Gaussian model always has the ideal capacity which is hard to be achieved in the real system. We can always find an equivalent Vandermonde model for the measured massive MIMO channel by our real experimental data, when their capacities are very close under optimal pair of $\alpha$ and $d$. 

In Table.~\ref{mme_tab}, the parameters pair ($d$, $\alpha$) with values of ($1.25\lambda /2$,  $28$ degree) gives the minimum MME value. This optimal parameters pair are very close to that in the real deployment ($20cm$, $26$ degree). 

%In Fig.~\ref{cap_vand} 
%the corresponding $\alpha$ is $35 ^\circ$ which is bigger than the real value $26 ^\circ$ in the deployed testbed. This is reasonable as the experiment is performed
%in the indoor environment. The signal multi-path scattering expands the equivalent range for angles of arrival .

%\begin{figure}
%\centering
%\includegraphics[width=3.6in]{figs/cap_vand.png}
%\caption{Comparison of measured channel with Vandermonde Channel Model by adjusting parameter $\alpha$ }
%\label{cap_vand}
%\end{figure}

\begin{table}[h!]
  \begin{center}
    \begin{tabular}{| l | l |l | l | l | l | l| l |}
    \hline
    $\alpha$ & MME-$d_1$ & $\alpha$ & MME-$d_2$ & $\alpha$ & MME-$d_3$ & $\alpha$ & MME-$d_4$ \\
    \bf{35} & 40.7338 & 35 & 86.2465 & 30 & 52.0346 & 30 & 77.9740\\
    30 & 98.9845 & \bf{32} & 39.7652 & 29 & 39.7046 & \bf{28} & 37.9838\\
		26 & 189.545 & 30 & 51.8876 & 26 & 66.8592 & 26 & 50.9662\\
    \hline
    \end{tabular}
  \end{center}
  \caption{Minimum mean error for different pair of $d$ and $\alpha$.}
	\label{mme_tab}
\end{table}
\section{conclusion}

This paper presents the design and implementation of a novel SDR based massive MIMO testbed using TDMA protocol.
The MIMO channel matrix is measured through the experiment with the prototype of the testbed. The system capacity is computed from the measured channel matrix. The 
testbed is validated by the linearity of capacity with the increasing number of antennas. We also found that
the measured channel capacity agrees with that of the Vandermonde channel model with the optimal parameters: the range of the arrival angles and the base station antenna distance which are very close to those in experimental deployment.  It is the first time the feasibility of the Vandermonde channel model is shown in the real experimental result. Based on our work, we recommend the Vandermonde channel model to be used in future theorectical research, as it is more realistic than the current widely used Gaussian model.

    \bibliographystyle{ieeetr}   

 \begin {small}
 
 \bibliography{bible/bibmimo}

\end {small}

\end{document}